\documentclass[conference]{IEEEtran}
\IEEEoverridecommandlockouts
\usepackage{cite}
\usepackage{amsmath,amssymb,amsfonts}
\usepackage{algorithmic}
\usepackage{graphicx}
\usepackage{textcomp}
\usepackage{xcolor}
\usepackage{booktabs} % for better horizontal rules in tables
\usepackage{multirow} % for multi-row cells

\def\BibTeX{{\rm B\kern-.05em{\sc i\kern-.025em b}\kern-.08em
    T\kern-.1667em\lower.7ex\hbox{E}\kern-.125emX}}
\begin{document}

\title{Correlation of Software-in-the-Loop Simulation with Physical Testing for Autonomous Driving}

\author{
    \IEEEauthorblockN{Zhennan Fei\IEEEauthorrefmark{1}, Mikael Andersson\IEEEauthorrefmark{1} and Andreas Tingberg\IEEEauthorrefmark{1}}
    \IEEEauthorblockA{\IEEEauthorrefmark{1}Volvo Cars\\
    Email: zhennan.fei@volvocars.com, andmika@gmail.com, andreas.tingberg@volvocars.com}
}

\maketitle

\begin{abstract}
Software-in-the-loop (SIL) simulation is a widely used method for the rapid development and testing of autonomous vehicles because of its flexibility and efficiency. This paper presents a case study on the validation of an in-house developed SIL simulation toolchain. The presented validation process involves the design and execution of a set of representative scenarios on the test track. To align the test track runs with the SIL simulations, a synchronization approach is proposed, which includes refining the scenarios by fine-tuning the parameters based on data obtained from vehicle testing. The paper also discusses two metrics used for evaluating the correlation between the SIL simulations and the vehicle testing logs. Preliminary results are presented to demonstrate the effectiveness of the proposed validation process. 
\end{abstract}

\begin{IEEEkeywords}
Software-in-the-loop simulation, vehicle testing, validation, correlation. 
\end{IEEEkeywords}

\section{Introduction}
% SIL
Autonomous vehicles (AV) have emerged as a promising technological advancement, revolutionizing the way we envision transportation in the modern era. With their potential to reduce accidents, increase traffic efficiency, and enhance accessibility, AV have garnered significant attention from researchers, policymakers, and industries alike. Despite the potential benefits, ensuring the safety and reliability of AV presents a formidable challenge. Traditional testing and verification methods, which rely on physical vehicle prototypes and real-world driving, are time-consuming and costly. Furthermore, due to the complexity of autonomous systems and the multitude of possible scenarios they must navigate, it becomes impractical and even hazardous to solely rely on physical testing for comprehensive safety assurance. 

Simulation plays a crucial role in the development, testing and validation of AV. It allows for the virtual representation of real-world scenarios, providing a controlled environment to test and evaluate autonomous driving system (ADS) components. Simulation encompasses different methodologies such as software-in-the-loop (SIL), hardware-in-the-loop (HIL) and vehicle-in-th-loop (VIL) with each serving a specific purpose in the development and testing life-cycle with the corresponding combination of hybrid simulation/real-hardware testing configuration \cite{huang_2016_av_testing_review}. This paper focuses on SIL but the presented validation method can possibly be applied to other approaches.

Software-in-the-loop (SIL) simulation refers to the simulation of the integrated software or its components and its operation in the real world utilizing simulation tools \cite{meng2019simulation}. SIL integrates the ADS under evaluation into a virtual environment, where simulated sensor inputs are processed for perception and planning. Control commands generated by the ADS determine the actions of the AV in the virtual environment, which simulates the AV's dynamics. The SIL simulation operates in a closed loop manner, with the virtual environment continuously providing sensor inputs to the ADS, which then generates control commands.

While SIL simulation enables evaluating the safety and performance of an ADS under a wide range of scenarios, including rare and hazardous situations, validating the tool-chain is a prerequisite and a vital step to establish confidence in the simulation's ability to replicate the behavior of the ADS in those scenarios. Moreover, as virtual testing such as SIL simulation is gradually becoming part of the certification process of future ADS \cite{Galassi2020certification}, the validation of simulation toolchain is essential for regulatory compliance. For instance, to meet the legal requirements of UN-R157 \cite{ALKS} for a SAE Level 3 \cite{SAE} autonomous driving system, it is necessary to establish correlation between the simulation toolchain and physical testing within the specific application domain.

The study detailed in \cite{Dona2022ADSValidationReview} offers a comprehensive exploration of contemporary advancements in validating toolchains for Autonomous Driving Systems (ADS) through virtual testing. 
Utilizing computational tools and diverse validation methodologies, the authors categorize the surveyed approaches into two main types. Firstly, there are "integrated-level system" validation strategies, where the primary emphasis is on assessing the overall performance of the entire virtual system. Secondly, there are "submodels-based" approaches, which involve breaking down the simulation pipeline into functional submodules such as the sensor model, vehicle model, and world model. These submodels are then individually validated step by step. While recognizing the importance of separately analyzing each virtual component within the testing pipeline, our paper specifically focuses on integrated-level validation. For an in-depth exploration of submodels-based approaches, we direct readers to \cite{Dona2022ADSValidationReview}.

In the integrated system validation, a reference maneuver is first defined, followed by the tuning of the considered simulation environment to reproduce the driving task virtually. Based on a list of Key Performance Indicators (KPIs) which are representative of the whole close-loop simulation environment, the validation is carried out accordingly. Several proposals falling under this category are presented in \cite{Riedmaier2018Validation, Dona2022ADSValidationExample}
and \cite{Riedmaier2020Validation}. 
In \cite{Riedmaier2018Validation}, a method for validating the Vehicle HIL (VeHIL) and SIL approaches compared to real driving tests is introduced and demonstrated in the use case of an self-developed Automated Longitudinal Driving (ALD) function. Starting with the proving ground test execution, the methods adopts precise measurement systems to provide the necessary accurate inputs to the ALD function assuming having a white box model while bypassing the actual environment sensor. The measured data is then transferred to the simulation environment to re-simulate the exact testing conditions from the proving ground. A number of KPIs are suggested for analyzing the correlation of the two simulation approaches with the  test track testing. 
In \cite{Dona2022ADSValidationExample}, a validation method similar to that of \cite{Riedmaier2018Validation} is presented where a VeHIL using camera simulation rather than signal injection is investigated in terms of the achievable fidelity level for increasing complexity driving scenarios. In addition to traditional discrepancy analysis, \cite{Dona2022ADSValidationExample} is complemented by adopting several computational tools for validating the VeHIL setup. On the other hand, the lateral dynamics of the vehicle is not part of the validation procedure due to the steering limitations.

In the light of the aforementioned work exploring various methodologies for validating virtual testing toolchains in the context of Autonomous Driving Systems (ADS), this paper contributes a unique case study centered around the validation of an in-house developed Software-in-the-Loop (SIL) simulation toolchain named CSPAS (Compiled Simulation Platform for Active Safety). The toolchain is specifically used during development and verification of safety requirements of Volvo Cars' autonomous driving solutions. The primary contributions of this paper are as follows: 
\begin{itemize}
    \item We introduce a comprehensive method and process for validating the considered SIL simulation toolchain, which is based on a representative subset of scenarios within the application domain.
    \item We propose an approach for effectively synchronizing test track runs. The approach further facilitates the fine-tuning of scenario parameters so that the set speed state for the ADS-vehicle and the trajectories of the target vehicle as those in the test track can be achieved in the SIL simulations.
    \item We discuss the metrics used to evaluate the correlation between the SIL simulations and the corresponding physical tests, and present some preliminary results. 
\end{itemize} 

The subsequent sections of the paper are organized as follows: Section~\ref{sec:validation_method_process} provides an overview of the validation method and process employed throughout this study. The design of the scenarios is detailed in Section~\ref{sec:scenario_design}, while Section~\ref{sec:vehicle_testing} delves into the test track testing setup used to execute these scenarios. Section~\ref{sec:synchronization_and_tuning} presents the approach utilized for the test track data synchronization, along with parameter tuning and simulating the refined scenarios. Section~\ref{sec:correlation_analysis} focuses on the discussion of two metrics employed for evaluating the correlation between SIL simulations and corresponding physical tests. Finally, the paper concludes in Section~\ref{sec:conclusion}, summarizing the contributions made by this study.

\section{Validation method and process}
\label{sec:validation_method_process}
To establish correlation with the real world, the SIL simulation results are compared against real world vehicle testing.
%As shown in Fig~\ref{fig:general_process}, this 
This involves conducting parallel testing, where the same version of the ADS is deployed both in the simulated environment and on physical vehicles. By performing an identical set of scenarios, the behavior, decisions and performance of the ADS in both the simulation and real-world scenarios are compared and evaluated to determine the correlations. 

% \begin{figure}[h!]
%     \centering
%     \includegraphics[width=0.3\textwidth]% {figs/general_validation.png}
%     \caption{A general process for validating SIL toolchain.}
%     \label{fig:general_process}
% \end{figure}

% \subsection{SIL Simulation Environment}
The SIL simulation toolchain being validated is an in-house developed platform called CSPAS, which is used to perform large scale scenario-based simulations to evaluate decision \& planning and control modules of ADS.

\begin{figure}[h!]
    \centering
    \includegraphics[width=0.475\textwidth]{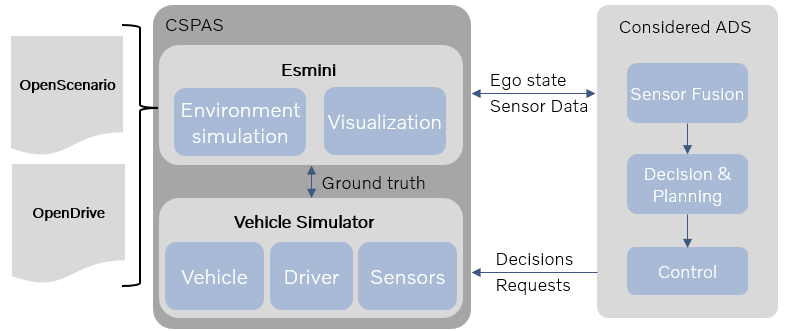}
    \caption{SIL simulation environment with CSPAS.}
    \label{fig:sim_env}
\end{figure}

The SIL simulation environment, featuring the CSPAS toolchain, is presented in Fig~\ref{fig:sim_env}. The architecture encompasses a vehicle simulator and the esmini open-source scenario player \cite{esmini}. The vehicle simulator is responsible for simulating the motion of the ADS-vehicle and incorporates sensor models to monitor the environment. It also includes a component for human driving, referred to as Driver, which is not directly relevant to the scope of this paper. The Environment simulation module within esmini handles the simulation of traffic objects, such as vehicles and vulnerable road users, adhering to the OpenSCENARIO \cite{asam-openscenario} and OpenDRIVE \cite{asam-opendrive} formats for representing scenarios and road infrastructure, respectively. To provide information about the surrounding actors and road environment in the vicinity of the simulated (AV), esmini is connected to the vehicle simulator, allowing for ground truth data exchange in the format of OSI (Open Simulation Interface)  \cite{asam-osi}. Finally, the visualization capability of esmini enables the visualization of scenarios.

\begin{figure}[h!]
    \centering
    \includegraphics[width=0.48\textwidth]{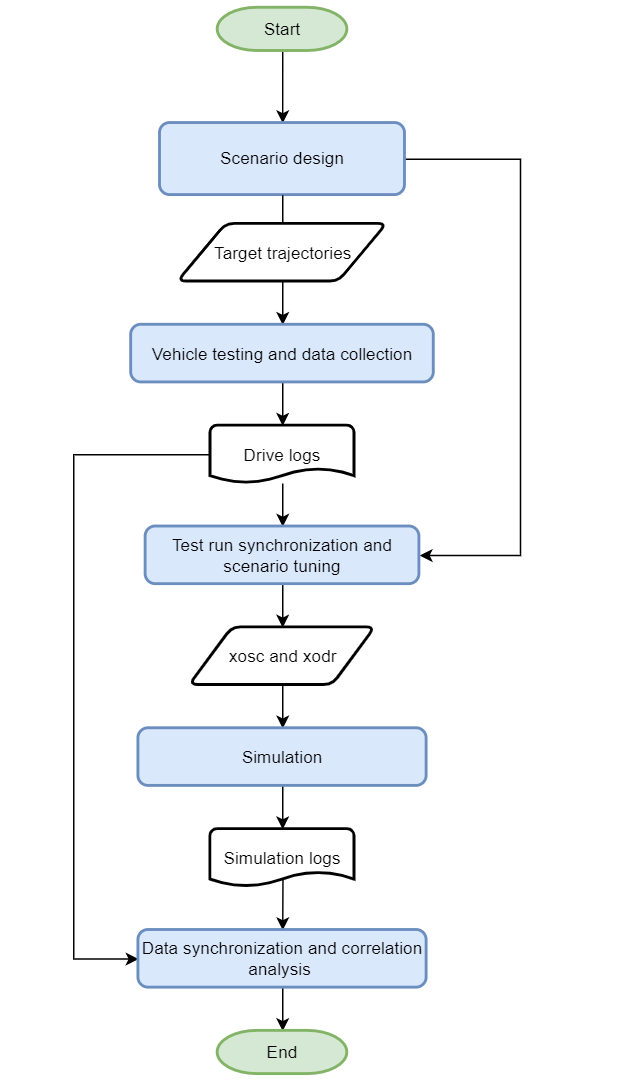}
    \caption{The proposed validation process.}
    \label{fig:sil-process}
\end{figure}

\begin{table*}[ht]
\centering
  \caption{Selection of designed scenarios}
  \label{tab:scenarioparameters}
\resizebox{18cm}{!} 
{ 
\begin{tabular}{lcccccl}\toprule
\multirow{2}{*}{Scenario} & \multicolumn{6}{c}{Parameters}
\\\cmidrule(lr){2-7}
           & ADS init speed ($km/h$)  & Target init speed ($km/h$) & Target decel ($m/s^2)$ & Trigger distance ($m$)  & Event duration ($s$) & Curvature ($m^{-1}$)\\\midrule
stationary-target & 70 & 0 & n/a & n/a & n/a & 0\\
cut-in & 30 & 20 & 1 & 60 & 10 & 0\\
cut-out & 55 & 55 & n/a & 40 & 4 & 1/140\\
\bottomrule
\end{tabular} }
\end{table*}

While SIL simulations with the CSPAS toolchain offers benefits for the rapid development and reduces the efforts for traditional vehicle testing, it does come with several simplifications that need to be considered. First, at the time of performing this case study, the vehicle simulator adopts ideal sensor models, which are not intended for modeling the sensors but rather reporting objects and road elements in the simulated scenarios. These sensor models include the respective field-of-view (FOV) and occlusion effects which occur when an object occludes another. 

Additionally, the sensor data that is generated by sensor models is consumed by the subsequence sensor fusion model which is an emulation of the sensor fusion component deployed in the test vehicle. The sensor signals reported by respective sensor models are fused within the sensor fusion model to generate sensor fusion output data, for example the tracked object list, which is then used by downstream components for decision-making, trajectory planning, and vehicle control in response to the traffic situation. 

Last but not at the least, the repeatability on test track is far from perfect. Besides environmental
factors such as road conditions, surface friction and visibility, it can be the case that the target
vehicle may experience difficulty following the exact trajectory of the predefined scenario.
%, particularly at low speeds, due to inaccuracies in torque and speed estimation
% which makes braking and propulsion control more challenging.

To mitigate the impact of these discrepancies and ensure an unbiased comparison, we introduce a validation process through a correlation study, illustrated in Fig~\ref{fig:sil-process}. The validation process, which will be elaborated in subsequent sections, aims to minimize the effects arising from these simplifications while establishing a reliable correlation between the SIL simulation and real-world testing.

\section{Scenario design}
\label{sec:scenario_design}
On the base of \cite{ALKS}, we parameterize the three scenario classes, namely "stationary target", "cut-in" and "cut-out\footnote{In the context of ALKS, the cut-out scenario involves exposing a secondary target, requiring the ADS-vehicle to perform deceleration.}" in terms of the following parameters:

% To be able to argue for the validity of a simulation toolchain such as CSPAS, the test scenarios should be designed to cover its application domain. Accordingly, the scenarios selected for this validation study are derived from the Operational Design Domain (ODD) of the ADS under consideration. While various scenario classes are taken into account, this paper focuses on the design of cut-in scenarios, which is summarized as follows.

% \begin{enumerate}
%    \item The ADS-vehicle and target vehicle drive in different lanes with different initial speeds.
%    \item At a certain distance between the two vehicles, the target makes a lane change in front of the ADS-vehicle.
%    \item When the lane change is done, the target starts to decelerate.
% \end{enumerate}

% This specific scenario can be parameterized as follows: 

\begin{itemize}
    \item The initial speeds of the ADS-vehicle.
    \item Both the initial speed and acceleration of the target vehicle.
    \item The relative distance between the ADS-vehicle and the target vehicle, which triggers the maneuver of the cut-in and cut-out vehicle.
    \item The duration for the maneuver of the target, i.e., the time it takes for the cut-in or cut-out vehicle to change lanes to/from the host lane of the ADS-vehicle.
    \item The curvature of the road.
\end{itemize}

% For the cut-in example a set of parameters can be listed to define the example scenarios, seen in Table \ref{tab:scenarioparameters}, where the parameters are ranging from easy to challenging where emergency actuation might be needed.  It is important to note that the target vehicle will not adapt its motion to the ADS-vehicle but follows the pre-defined trajectory or remains stationary. The ADS-vehicle will constantly adjust its position based on the detected targets and lane markers.

Table~\ref{tab:scenarioparameters} presents three selected scenarios parameterized based on the previously mentioned criteria. During the correlation study, a number of variations have been carried out. However, it is worth mentioning that the emphasis here is not on the specific traffic scenarios themselves, but rather on their role as tools for eliciting particular responses from the ADS. As a result, these scenarios have been
designed with the specific intent of provoking distinct triggering moments. For instance, the
"stationary target" scenarios serve as a test of the ADS-vehicle's capacity to respond to an object
detected at a considerable distance. Those scenarios are generally considered to give the ADS-vehicle a long planning horizon.

In contrast, the "cut-in" scenarios are tailored to induce a range of responses from the ADS, spanning from nominal deceleration to emergency braking, thereby evaluating the ADS-vehicle's ability to react with varying levels of urgency. Those scenarios are used to create an urgent
situation and stress test the avoidance predictability of the autonomous driving function.

Lastly, the "cut-out" scenarios are structured to examine how the ADS-vehicle responds to a transition to a secondary target, often prompting abrupt deceleration. For those scenarios, the lead vehicle exposes a previously occluded
secondary target during the cut-out maneuver. The measure is to evaluate the correlation of the deceleration profile when the ADS-vehicle reacts to the previously occluded stationary target. Note that for the "cut-out" scenarios, it is the exposure time to the secondary target that matters. The
exposure time is indirectly defined by the combination of the parameter values of the cut-out vehicle.

The scenarios were designed using the Python package scenariogeneration \cite{scenariogeneration}. This package provides a user-friendly approach for creating scenarios by inheriting a pre-defined Python class (ScenarioGenerator) and realizing specific abstract methods that take as input the aforementioned scenario parameters. Utilizing these parameters and the different modules in scenariogeneration, all the necessary OpenDRIVE (.xodr) and OpenSCENARIO (.xosc) files can be generated.

The generated .xosc and .xodr scenario files are passed as input to esmini, which executes the scenarios, both for visual inspection as well as the ground truth of the whole scenario. One of the outputs from esmini is an Open Simulation Interface (OSI) file that contains ground truth information for the scenario. From the generated ground truth data, scripts were developed to extract the trajectories of the targets in terms of time, and x-, and y-coordinates as well as the bumper-to-bumper distance that triggers the maneuver of the target. This data is saved in a textual .pmc file, which can be used to control the target in the subsequent data collection step.

\section{Vehicle Testing and Data Collection}
\label{sec:vehicle_testing}

%\subsection{Test track environment}
The vehicle testing of the designed scenarios in the case study was conducted at AstaZero \cite{astazero}, a test facility containing various track areas specifically designed for performing different types of traffic scenarios. 

\begin{figure}[h!]
    \centering
    \includegraphics[width=0.4\textwidth]{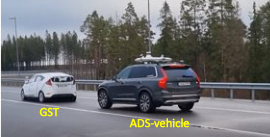}
    \caption{The AV under testing and the GST vehicle on AstaZero.}
    \label{fig:gts-av}
\end{figure}

Figure~\ref{fig:gts-av} illustrates the test object, the ADS-vehicle, and a guided soft target (GST) vehicle, which serves as the cut-in vehicle for the cut-in scenarios and the stationary vehicle in stationary target and cut-out scenarios according to Table \ref{tab:scenarioparameters}.

The GST vehicle features a low platform with a soft target mounted on top. Its behavior is governed by the reference trajectory provided by the previously generated .pmc file. The GST vehicle utilizes feedback controllers to regulate its speed and follow the reference path.

To facilitate data collection and monitoring, a driving robot is installed in the ADS-vehicle. The robot's primary function is to log the positions of the ADS-vehicle and track the positions of the target vehicle through wireless communication. The robot does not engage in steering or speed control maneuvers during the tests.

During the vehicle testing phase, the lane width was fixed at 3.5 meters. For the curved road scenarios, the ADS-vehicle was expected to steer right while remaining in the center of the lane. % The target vehicle executed cut-in maneuvers from the inner curve and steered left toward the host lane.

The output of the test track testing consists of two streams of data logs:
\begin{itemize}
    \item Drive logs: These logs consist of signals outputted by the software components of the considered ADS. Upon decoding, the logs are stored in HDF5 file format. In this paper, the signals from the drive logs are used to conduct the correlation analysis.
    \item Position logs: These logs record the high precision position data collected by the robot, relative to a local coordinate system.
\end{itemize}

\section{Data synchronization and scenario tuning}
\label{sec:synchronization_and_tuning}

\subsection{Data synchronization}
Due to environmental factors and manual driving to initiate scenarios under testing leading to
inaccuracies in vehicle positioning, achieving a high degree of repeatability in physical track tests
poses a significant challenge. For the effective correlation of logged scenarios, it is essential to
initialize all vehicles in identical positions across every repetition. Given the insufficiency of manual
setup and triggering, a synchronization process, designed to align repetitions for each scenario,
becomes necessary.

In the validation of the CSPAS simulation tool, each scenario delineated in Table \ref{tab:scenarioparameters} was executed $5$
times on the test track to account for potential variations. Those which were executed and annotated successfully
were then selected for the analysis. In order to align these repetitions and derive statistics on signal
correlations with simulations, we have developed a synchronization mechanism, which is outlined in
the subsequent steps.

\begin{enumerate}
    \item Determine the relative distance, denoted by $S_{d}$, at which the ADS-vehicle initiates its response to the target vehicle for each repetition of the same scenario.  This process varies depending on the scenario class:
    \begin{itemize}
        \item In stationary target scenarios, we define $S_d$ as the distance at which the target enters the ADS-vehicle’s field of view (FOV).
        \item For cut-in scenarios, $S_d$ is calculated as the distance to the cut-in vehicle when it enters the same lane as the ADS-vehicle.
        \item In cut-out scenarios, $S_d$ represents the distance to the occluded vehicle when it becomes detected by the ADS-vehicle.
    \end{itemize}
    \item Among the relative distances identified in the previous step across all iterations for each scenario, we select the shortest distance, denoted as $S_{min}$. Then we pinpoint the corresponding timestamp, labeled as $t_{min}$, from the drive log.
    \item We trace back from the timestamp $t_{min}$, as identified in step 2, within the drive log. The objective here is to pinpoint an earlier timestamp at which the ADS-vehicle’s relative distance to the target, denoted as $S_{sync}$, falls within the range $(S_{at},S_{min})$, for example according to Eq~\ref{eq:sync_dist}. The parameter $S_{at}$ signifies the minimal distance to the target among all repetitions when the beginning of the scenario is annotated (by the test engineer in the passenger seat of the ADS-vehicle), implying ADS has been activated and the ADS-vehicle on the test track has also reached a stable, steady-state condition.
    \begin{equation}\label{eq:sync_dist}
        S_{sync} = \frac{S_{at}-S_{min}}{2}
    \end{equation}
    \item To synchronize data from other repetitions, we employ $S_{sync}$ as a reference point. This involves calculating the integral of velocity alongside corresponding timestamps.
\end{enumerate}

\begin{figure}[h!]
    \centering
    \includegraphics[width=0.35\textwidth]{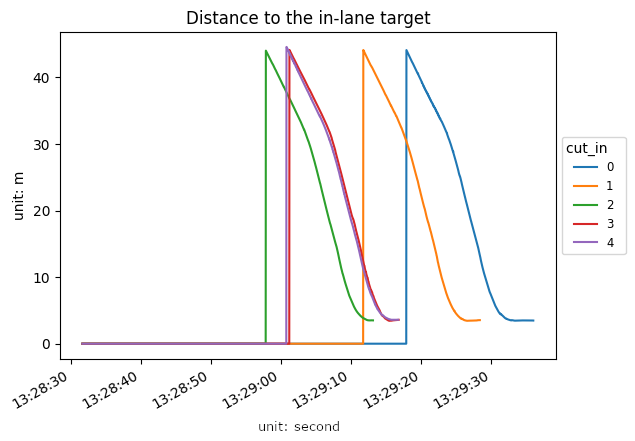}
    \caption{Unsynchronized plot for the relative distance to the target for the cut-in scenario in Table~\ref{tab:scenarioparameters}.}
    \label{fig:inlan_tgt_dist_cutin_1_unsynch}
\end{figure}

\begin{figure}[h!]
    \centering
    \includegraphics[width=0.35\textwidth]{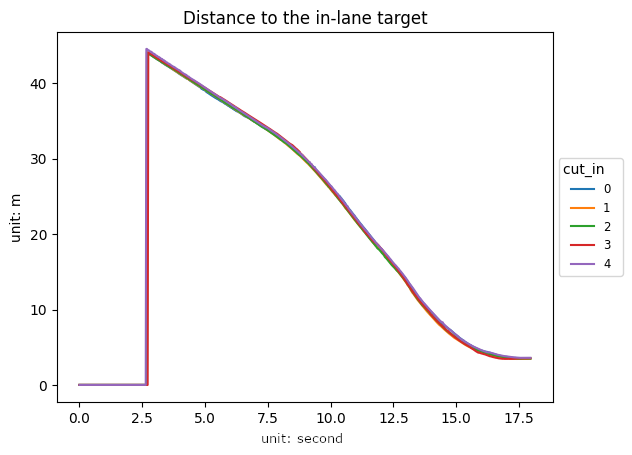}
    \caption{Synchronized plot for the relative distance to the target for the cut-in scenario in Table~\ref{tab:scenarioparameters}.}
    \label{fig:inlan_tgt_dist_cutin_1_synch}
\end{figure}

As an example, Fig.~\ref{fig:inlan_tgt_dist_cutin_1_unsynch} depicts the plot of the signal representing the remaining relative distance to the (in-lane) target extracted directly from the corresponding drive logs based on the annotations of repetitions. By utilizing the outlined method, Fig.~\ref{fig:inlan_tgt_dist_cutin_1_synch} depicts the plot of this signal which are synchronized.

\subsection{Scenario tuning}
The aforementioned synchronization procedure not only allows for the alignment of recorded
scenario repetitions from the test track, despite any inaccuracies, but also provides a method for
adjusting the keys parameters to ensure identical scenario setup in the simulation.

\emph{Remark.} The objectives of scenario tuning are twofold: (i) to achieve a consistent steady-state for the
ADS-vehicle in the simulation, mirroring the ADS-vehicle’s state on the test track before reacting to
the target, and (ii) to guarantee that the target’s trajectories in both the simulation and test track are
in alignment. Consequently, adjustments made to the scenario parameters do not affect the
subsequent responses once the target is detected.

The following describes the scenario parameters and the process of adjusting their values in the simulation based on the synchronized repetitions:

\begin{itemize}
    \item  The ADS-vehicle set speed is fine-tuned by averaging the initial speeds of the test object across the synchronized repetitions.
    \item  The maximal range of the ideal sensor, which is applicable to scenario such as stationary targets and cut-out, is calibrated to correspond with the performance of the test object during test track tests.
    % \item The speed profile of the target vehicle is fine-tuned to align with the average speed and acceleration of the test object among the synchronized repetitions.  
\end{itemize}

By employing the polished scenarios with carefully adjusted parameters, the associated OpenSCENARIO and OpenDRIVE files are generated for simulation purposes.

\subsection{Simulation}

The CSPAS simulation toolchain, powered by esmini, takes previously generated XML-based xosc and xodr files as inputs. The “World Engine” of esmini is used to interpret the input OpenSCENARIO and OpenDRIVE files and create the environment. The CSPAS includes a vehicle simulator that models the characteristics of the ADS-vehicle.

By applying the same synchronization procedure discussed previously, the simulation results are synchronized with the repetitions of the respective scenario conducted during test track testing.

With the logs from both test track trials and simulation synchronized, we are now well-positioned to proceed with the correlation analysis.

\section{Correlation Analysis}
\label{sec:correlation_analysis}

This section presents the selection of representative signals employed for comparing the simulation outcomes with the test track evaluation results, the visualizations of these signals, and the results derived from applying evaluation metrics to the three scenarios in Table~\ref{tab:scenarioparameters}.

Specifically, the following signals are used for the correlation analysis: (i) $s_{dist}$: the relative distance between the ADS-vehicle and the target after it is detected as a potential threat. As our purpose is to assess how the ADS-vehicle
responds to the target in both the test track testing and the simulation, it is essential to synchronize
the initial value of $s_{dist}$ in the simulation with the values obtained from the test track repetitions. This
synchronization is achieved by adjusting the sensor models' range to ensure that the initial $s_{dist}$ in the simulation falls within the range of values observed in the repeated scenarios on the test track. Note that this relative distance, after the target is detected, is no longer a sensing performance but indicates how the ADS-vehicle reacts to the target which is either stationary
or brakes. From the toolchain validation point of view, comparing this KPI between the simulation and test track testing is crucial. 
(ii) $v_{lon}$: the longitudinal velocity of the ADS-vehicle. Following the same rationale, the initial value of $v_{lon}$ representing the longitudinal speed when the
target is first identified as a potential threat, is harmonized with the values derived from the test track
repetitions. Similar to $s_{dist}$, the longitudinal speed $v_{lon}$ can indicate how the vehicle reacts to the target which is either stationary or brakes. (iii) $a_{lon}$: the longitudinal acceleration of the ADS-vehicle and (iv) $a_{lat}$: the lateral acceleration of the AD-vehicle for the scenarios conducted on the curved road. It is worth noting that the simulation and test track testing environments, as well as
discrepancies between the vehicle dynamics model and the actual vehicle, can cause the longitudinal acceleration ($a_{lon}$) and the lateral acceleration ($a_{lat}$) to differ significantly in some cases. Nevertheless, it is still important
to ensure that these signals are adequately correlated.

\begin{figure}[h!]
    \centering
    \includegraphics[width=0.38\textwidth]{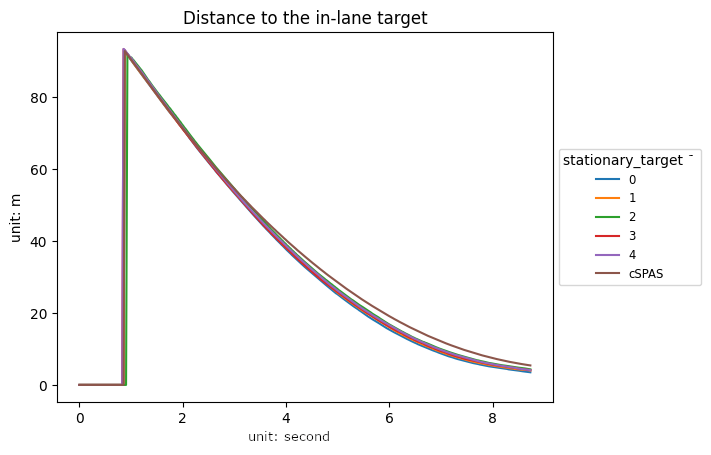}
    \caption{Signal $s_{dist}$ for the stationary target scenario in Table~\ref{tab:scenarioparameters}.}
    \label{fig:inlan_dist_st_1}
\end{figure}

\begin{figure}[h!]
    \centering
    \includegraphics[width=0.38\textwidth]{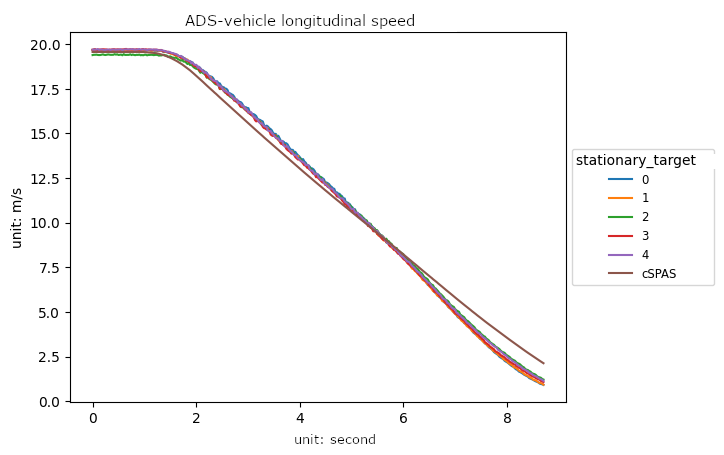}
    \caption{Signal $v_{lon}$ for the stationary target scenario in Table~\ref{tab:scenarioparameters}.}
    \label{fig:ego_spd_st_1}
\end{figure}

\begin{figure}[h!]
    \centering
    \includegraphics[width=0.38\textwidth]{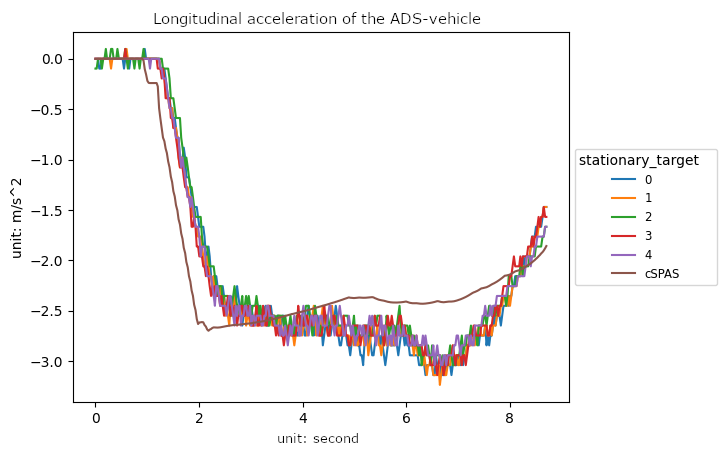}
    \caption{Signal $a_{lon}$ for the stationary target scenario in Table~\ref{tab:scenarioparameters}.}
    \label{fig:ego_acc_st_1}
\end{figure}

\begin{figure}[h!]
    \centering
    \includegraphics[width=0.35\textwidth]{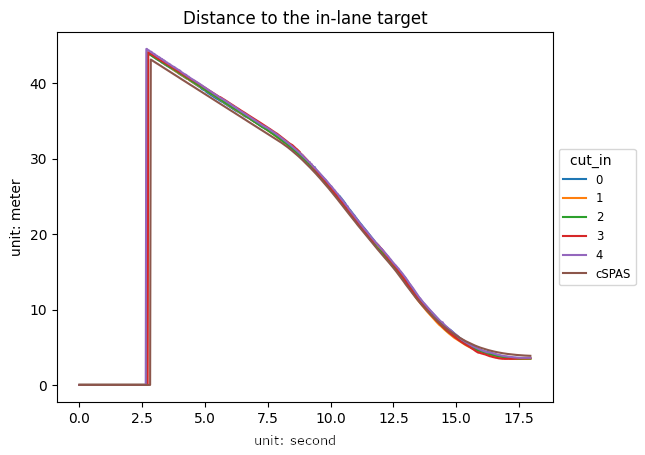}
    \caption{Signal $s_{dist}$ for the cut-in scenario in Table~\ref{tab:scenarioparameters}.}
    \label{fig:inlan_dist_cutin_1}
\end{figure}

\begin{figure}[h!]
    \centering
    \includegraphics[width=0.35\textwidth]{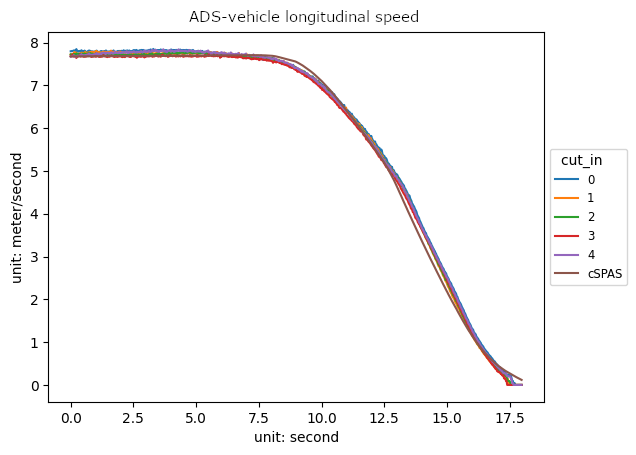}
    \caption{Signal $v_{lon}$ for the cut-in scenario in Table~\ref{tab:scenarioparameters}.}
    \label{fig:ego_spd_cutin_1}
\end{figure}

\begin{figure}[h!]
    \centering
    \includegraphics[width=0.35\textwidth]{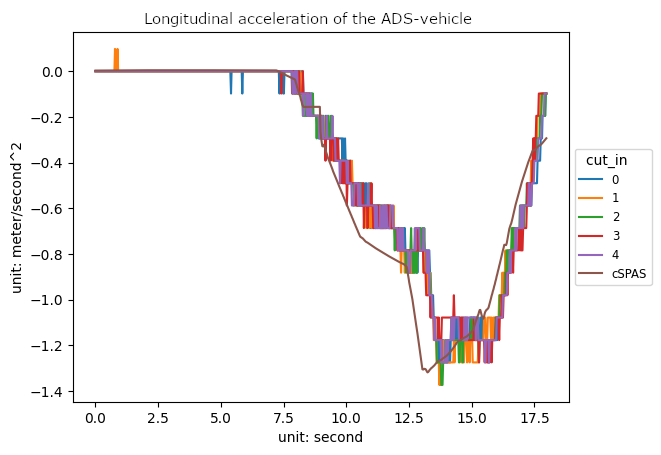}
    \caption{Signal $a_{lon}$ for the cut-in scenario in Table~\ref{tab:scenarioparameters}.}
    \label{fig:ego_acc_cutin_1}
\end{figure}

\begin{figure}[h!]
    \centering
    \includegraphics[width=0.35\textwidth]{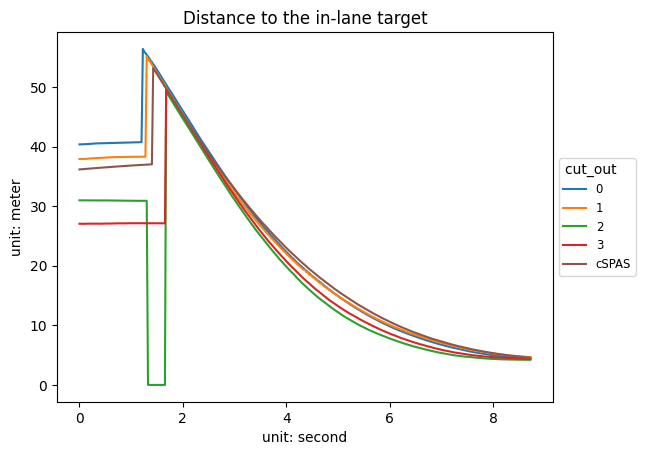}
    \caption{Signal $s_{dist}$ for the cut-out scenario in Table~\ref{tab:scenarioparameters}.}
    \label{fig:inlan_dist_cutout_2}
\end{figure}

\begin{figure}[h!]
    \centering
    \includegraphics[width=0.35\textwidth]{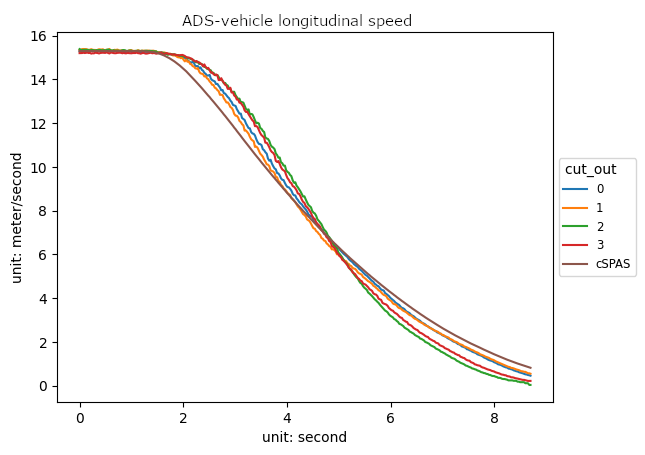}
    \caption{Signal $v_{lon}$ for the cut-out scenario in Table~\ref{tab:scenarioparameters}.}
    \label{fig:ego_spd_cutout_2}
\end{figure}

\begin{figure}[h!]
    \centering
    \includegraphics[width=0.35\textwidth]{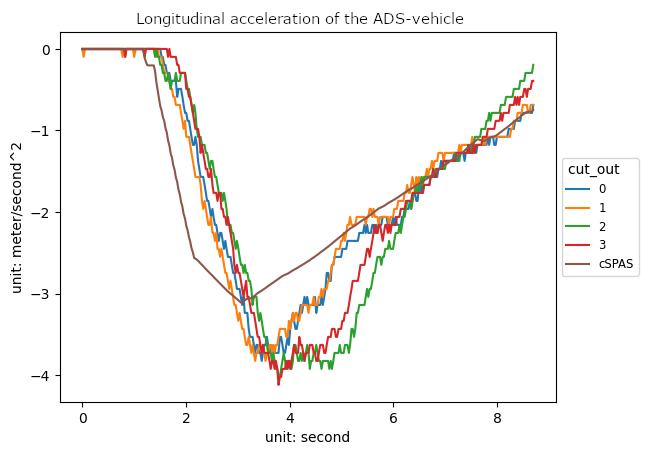}
    \caption{Signal $a_{lon}$ for the cut-out scenario in Table~\ref{tab:scenarioparameters}.}
    \label{fig:ego_acc_cutout_2}
\end{figure}

\begin{figure}[h!]
    \centering
    \includegraphics[width=0.35\textwidth]{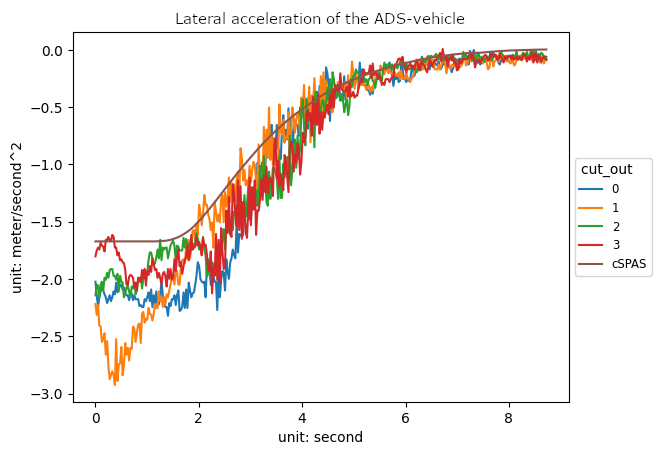}
    \caption{Signal $a_{lat}$ for the cut-out scenario in Table~\ref{tab:scenarioparameters}.}
    \label{fig:ego_acc_lateral_cutout_2}
\end{figure}

Fig.~\ref{fig:inlan_dist_st_1} -- Fig.~\ref{fig:ego_acc_lateral_cutout_2} depict respectively the plots of the aforementioned signals as KPIs for the scenarios stationary target, cut-in and cut-out listed in Table~\ref{tab:scenarioparameters}. Note that the plot the signal representing the lateral acceleration of the ADS-vehicle is only shown for the cut-out scenario as it is the only one among the chosen three which was conducted on the curved road.

% When validating the simulation toolchain from the complete feature point of view, it is essential to determine which signals are most relevant for establishing confidence in ADS's performance and its collision avoidance predictions. In this regard, the remaining distance to the target ($s_{dist}$) and the longitudinal velocity ($v_{lon}$) are the two most critical signals. 
%These signals should be well-correlated between simulation and test track testing, with deviations that are sufficiently small.

% However, it is worth noting that the simulation and test track testing environments, as well as discrepancies between the vehicle dynamics model and the actual vehicle, can cause %the acceleration request ($a_{lon}^{req}$) and 
% the actual acceleration ($a_{lon}$) to differ in some cases. Nevertheless, it is still important to ensure that these signals are adequately correlated.

\subsection{Correlation analysis}
Two evaluation metrics, namely Pearson correlation and Relative Root Mean Squared Error, are applied in the correlation analysis in the case study.

\textbf{Pearson Correlation} \cite{moore2018introduction, Riedmaier2018Validation} 
 is a metric employed to measure how the performed simulations are linearly related to the test track data. 
%As shown in Eq~(\ref{eq:cor-eff}), the Pearson's correlation coefficient of two variables $X$ and $Y$ in question is obtained by taking the ratio of their covariance of the two variables, normalized to the square root of their variances. 
The Pearson correlation coefficient is a number between -1 and 1, which expresses the degree that, on an average, two variables change correspondingly.

%\begin{equation}\label{eq:cor-eff}
%\rho = \frac{Covariance(X, Y)}{\sqrt{Variance(X)} \cdot \sqrt{Variance(Y)}}
%\end{equation}

Given a series of $n$ measurements for a signal from test track data, $M = \{m_1, \ldots, m_n\}$, and a series of $n$ measures for the same signal but from the simulation, $G = \{g_1, \ldots, g_n\}$, the sample correlation coefficient, $r$, defined in Eq~(\ref{eq:est-cor-eff}) can be used to estimate the Pearson's correlation coefficient between $M$ and $G$.

\begin{equation}\label{eq:est-cor-eff}
r = \frac{ \sum\limits_{i=1}^{n}(m_i-\bar{m})(g_i-\bar{g})}{\sqrt{\sum\limits_{i=1}^{n}(m_i-\bar{m})^2}\sqrt{\sum\limits_{i=1}^{n}(g_i-\bar{g})^2}},
\end{equation}
where $\bar{m}$ and $\bar{g}$ are the mean values for $M$ and $G$, respectively.
In addition, to evaluate if the correlation coefficients are statistically significant, their P-values are calculated and compared with the conventional threshold $5\%$. In theory, the P-value is the probability that one would have found the current result if the correlation coefficient were in fact zero (i.e., null hypothesis). If this probability is lower than the threshold, the calculated correlation coefficient is statistically significant. 

Given the P-values for the calculated coefficients $r$ for all scenarios being sufficiently small ($\ll0,05$), the following ranges of magnitude of $r$ indicates how two signals are positively correlated \cite{moore2018introduction}: (i) $0.7 - 1.0$ for high linear correlation, (ii) $0.5 - 0.7$ for moderate linear correlation, (iii) $0.3 - 0.5$ for low linear correlation and (iv) $0.1 - 0.3$ for weak linear correlation.

\textbf{Relative Root Mean Squared Error}. While the correlation coefficient indicates the magnitude of how two variables are related positively or negatively, it does not reveal how close the values of the same signal are in between the simulations and test track testing. 
To this end, a variant of Root Means Square Error (RMSE) \cite{Maupin2017ValidationMF, Dona2022ADSValidationReview}, Relative Root Mean Error (RRMSE) to evaluate their averaged difference. 
As shown in Eq~(\ref{eq:rrmse}), RRMSE is RMSE normalized by the root mean square (RMS) where each residual is scaled against the actual value. 

\begin{equation}\label{eq:rrmse}
RRMSE = \frac{RMSE}{RMS} = \frac{\sqrt{\frac{1}{n} \sum\limits_{i=1}^{n}(m_i - g_i)^2}}{\sqrt{\frac{1}{n}\sum\limits_{i=1}^n{(m_i)^2}}}.
\end{equation}

RRMSE expresses the error relatively or in a percentage form. Empirically, model accuracy is (i) excellent when RRMSE $< 10\%$, (ii) good when RRMSE is between $10\%$ and $20\%$, (iii) fair when RRMSE is between $20\%$ and $30\%$ and (iv) poor when RRMSE $> 30\%$.

\subsection{Results}
Table~\ref{tab:metrics_st_1}, \ref{tab:metrics_cutin_1} and \ref{tab:metrics_cutout_2} show the results of applying the metrics to the three aforementioned KPIs to indicate the correlation of the simulation of each scenario with the corresponding annotated repetitions on the test track along with their mean~\footnote{The ``mean'' repetition is calculated by averaging signal values of all synchronized repetitions at each timestamp.}. According to the Pearson correlation coefficient $r$, calculated according to Eq.~(\ref{eq:est-cor-eff}), the simulation exhibits a strong correlation with the test track testing, with the majority of r values exceeding $0.9$. On the other hand, the calculated values of RRMSE according to Eq.~(\ref{eq:rrmse}), which gauge the accuracy of signals between the simulation and the test track testing, appear to be more sensitive to specific scenarios and the signals employed for comparison. As indicated by the results, the signal $v_{lon}$ consistently demonstrates excellent accuracy, followed by good accuracy for $s_{dist}$ and varying between good to fair accuracy for acceleration-related signals. Notably, two repetitions of the challenging cut-out scenario yielded sub-optimal results on the test track, impacting the accuracy of acceleration-related signals in these instances.

\section{Conclusion}
The validation of a simulation toolchain is crucial to ensure its correlation with the outcomes of physical testing.
In this paper, we presented a validation study of the in-house developed simulation toolchain CSPAS. By following the proposed process, a set of representative scenarios are designed and conducted on the test track. In our case study, we introduce a procedure to synchronize scenario repetitions from test track as well a fine tuning and synchronization of the corresponding simulation scenarios to the scenarios realized in the vehicle tests.
In addition, we evaluated the correlation using metrics such as Pearson Correlation and RRMSE.

Overall, the study contributes to the validation and improvement of SIL simulation toolchains for autonomous vehicle by introducing a lightweight process and method for analyzing the correlation of KPIs between simulation and physical testing. 

As future work, advanced metrics will be explored to provide a comprehensive assessment of the simulation's performance. These advanced metrics may offer deeper insights into the correlation between the simulation and real-world testing, thereby enhancing our understanding of the simulation toolchain's effectiveness. 

% The findings enhance our confidence in the accuracy and reliability of the CSPAS platform, paving the way for further advancements in autonomous vehicle development and testing.

\begin{table}[ht]
\centering
  \caption{Metrics for stationary target (R: RRMSE)}
  \label{tab:metrics_st_1}
\begin{tabular}{lcccccl}\toprule
\multirow{2}{*}{Rep} & \multicolumn{2}{c}{$s_{dist}$} & \multicolumn{2}{c}{$v_{lon}$} & \multicolumn{2}{c}{$a_{lon}$}
\\\cmidrule(lr){2-3}\cmidrule(lr){4-5}\cmidrule(lr){6-7}
           & $r$  & $R~(\%)$ & $r$    & $R~(\%)$  & $r$ & $R~(\%)$\\\midrule
0 & 0.99 & 5.80  & 0.99 & 5.28 & 0.91 & 18.87 \\
1 & 0.99 & 4.87  & 0.99 & 4.99 & 0.93 & 17.04 \\
2 & 0.96 & 17.06 & 0.99 & 3.93 & 0.92 & 17.78 \\
3 & 0.98 & 12.71 & 0.99 & 4.56 & 0.94 & 15.53 \\
4 & 0.98 & 12.38 & 0.99 & 4.30 & 0.94 & 15.49 \\
mean & 0.99 & 7.35 & 0.99 & 4.58 & 0.91 & 14.70 \\\bottomrule
\end{tabular}
\end{table}

\begin{table}[ht]
\centering
  \caption{Metrics for cut-in (R: RRMSE)}
  \label{tab:metrics_cutin_1}
\begin{tabular}{lcccccl}\toprule
\multirow{2}{*}{Rep} & \multicolumn{2}{c}{$s_{dist}$} & \multicolumn{2}{c}{$v_{lon}$} & \multicolumn{2}{c}{$a_{lon}$}
\\\cmidrule(lr){2-3}\cmidrule(lr){4-5}\cmidrule(lr){6-7}
           & $r$  & $R~(\%)$ & $r$    & $R~(\%)$  & $r$ & $R~(\%)$\\\midrule
0 & 0.97 & 15.59 & 0.99 & 2.72 & 0.95 & 23.41 \\
1 & 0.97 & 15.45 & 0.99 & 1.92 & 0.96 & 20.69 \\
2 & 0.98 & 14.32 & 0.99 & 1.91 & 0.96 & 21.39 \\
3 & 0.98 & 12.87 & 0.99 & 2.07 & 0.95 & 22.83 \\
4 & 0.97 & 16.55 & 0.99 & 2.23 & 0.96 & 22.79 \\
mean & 0.98 & 14.20 & 0.99 & 2.07 & 0.96 & 21.31 \\
\bottomrule
\end{tabular}
\end{table}

\begin{table}[!h]
\centering
  \caption{Metrics for cut-out (R: RRMSE)}
  \label{tab:metrics_cutout_2}
\begin{tabular}{lcccccccl}\toprule
\multirow{2}{*}{Rep} & \multicolumn{2}{c}{$s_{dist}$} & \multicolumn{2}{c}{$v_{lon}$} & \multicolumn{2}{c}{$a_{lon}$} & \multicolumn{2}{c}{$a_{lat}$}
\\\cmidrule(lr){2-3}\cmidrule(lr){4-5}\cmidrule(lr){6-7}\cmidrule(lr){8-9}
           & $r$  & $R~(\%)$ & $r$ & $R~(\%)$  & $r$ & $R~(\%)$ & $r$ & $R~ (\%)$\\\midrule
0 & 0.98 & 11.28 & 0.99 & 4.37 & 0.90 & 24.66 & 0.98 & 27.08 \\
1 & 0.98 & 8.1 & 0.99 & 3.56 & 0.93 & 21.34 & 0.96 & 28.91\\
2 & 0.79 & 44.26 & 0.99 & 9.07 & 0.81 & 36.89 & 0.98 & 21.64\\
3 & 0.94 & 25.83 & 0.99 & 7.69 & 0.83 & 33.98 & 0.97 & 21.91 \\
mean & 0.97 & 14.08 & 0.99 & 6.05 & 0.88 & 28.42 & 0.99 & 21.83
\\\bottomrule
\end{tabular}
\end{table}

\label{sec:conclusion}

%\section*{Acknowledgment}
%The authors would like to express our gratitude to the colleagues at Volvo Cars for conducting the vehicle testing, annotating the scenarios on the test track and reviewing the draft of the paper. Special thanks go to Matthias Eng, Anders \"{O}dblom, Emil Knabe, Zijian Han and Magnus Larsson for having the fruitful discussions and proving the technical support. 

%*******************************************************************************
% BIBLIOGRAPHY
%*******************************************************************************
\bibliographystyle{IEEEtran}
\bibliography{references}

\end{document}